\documentclass[conference]{IEEEtran}
\IEEEoverridecommandlockouts
\usepackage{cite}
\usepackage{amsmath,amssymb,amsfonts}
\usepackage{algorithmic}
\usepackage{graphicx}
\usepackage{textcomp}
\usepackage{stfloats}
\usepackage{multirow}
\usepackage{xcolor}

\usepackage{makecell}

\def\BibTeX{{\rm B\kern-.05em{\sc i\kern-.025em b}\kern-.08em
    T\kern-.1667em\lower.7ex\hbox{E}\kern-.125emX}}
\begin{document}
\bibliographystyle{unsrt}

\title{\textbf{APEX}: \textbf{\underline{A}}ttention on \textbf{\underline{P}}ersonality based \textbf{\underline{E}}motion Re\textbf{\underline{X}}gnition Framework
}

\author{

\IEEEauthorblockN{Ruijie Fang}
\IEEEauthorblockA{\textit{ECE Department} \\
\textit{University of California, Davis }\\
CA, USA\\
rjfang@ucdavis.edu}

\and
\IEEEauthorblockN{Ruoyu Zhang}
\IEEEauthorblockA{\textit{ECE Department} \\
\textit{University of California, Davis}\\
CA, USA}

\and
\IEEEauthorblockN{Elahe Hosseini}
\IEEEauthorblockA{\textit{ECE Department} \\
\textit{University of California, Davis}\\
CA, USA}

\and
\IEEEauthorblockN{Chongzhou Fang}
\IEEEauthorblockA{\textit{ECE Department} \\
\textit{University of California, Davis}\\
CA, USA}

\and
\IEEEauthorblockN{Mahdi Eslaminehr}
\IEEEauthorblockA{\textit{Quandary Peak Research } \\
CA, USA \\
mahdi@quandarypeak.com}

\and
\IEEEauthorblockN{Setareh Rafatirad}
\IEEEauthorblockA{\textit{CS Department} \\
\textit{University of California, Davis}\\
CA, USA}
\and
\IEEEauthorblockN{Houman Homayoun}
\IEEEauthorblockA{\textit{ECE Department} \\
\textit{University of California, Davis}\\
CA, USA}
}

\maketitle

\begin{abstract}
Automated emotion recognition has applications in various fields, such as human-machine interaction, healthcare, security, education, and emotion-aware recommendation/feedback systems. Developing methods to analyze human emotions accurately is essential to enable such diverse applications. Multiple studies have been conducted to explore the possibility of using physiological signals and machine-learning techniques to evaluate human emotions. Furthermore, internal factors such as personality have been considered and involved in emotion recognition. However, integrating personality that is user specific within traditional machine-learning methods that use user-agnostic large data sets has become a critical problem. This study proposes the APEX: attention on personality-based emotion recognition framework, in which multiple weak classifiers are trained on physiological signals of each participant's data, and the classification results are reweighed based on the personality correlations between corresponding subjects and test subjects. Experiments have been conducted on the ASCERTAIN dataset, and the results show that the proposed framework outperforms existing studies.

\end{abstract}


\section{Introduction}
Emotion is a mental state caused by neurophysiological changes associated with thoughts, feelings, and behavioral responses, which play an essential role in human life \cite{dzedzickis2020human}. From the perspective of evolution, emotion is a product of natural selection as it helps humans survive. For example, fear causes humans to avoid harm, joy makes us repeat what works, sadness nudges us to ask for help, and the emotion of disgust makes us spit up accidentally eaten foreign objects. In modern society, emotion is essential as it directly influences people's activity and productivity. Positive emotions can promote coordination and organization, which is conducive to improving productivity. In contrast, negative emotions can make people feel bored, depressed, and dull and negatively affect people's creative thinking \cite{vanderlind2020understanding, fang2022atlas}.

Emotion has three distinct components: a subjective experience, external behaviors, and physiological arousal. Subjective experience is an individual's self-feeling of different emotional states \cite{zhang2020emotion, fang2022towards,hosseini2023emotion}. External behaviors of emotions, often called expressions, is the quantified form of actions of various parts of the body when an emotional state occurs, including facial expressions, gesture expressions, and intonation expressions \cite{fang2022pain,zhang2023short}. Physiological arousal refers to the physiological response to emotion, including a series of responses in the autonomic nervous system (ANS). Multiple studies have been conducted to automatically recognize emotion based on external behaviors or physiological signals. These methods exploit the connections between emotions, external behaviors, and physiological arousal to automatically recognize emotion and apply it in different applications e.g. human-machine interface, neuro-marketing, and healthcare \cite{shu2018review, zhang2023privee}.

Automated emotion recognition can be categorized into behavior-based and physiological signal-based approaches. The behavior-based approach captures various body activities, including facial expressions, body posture, gestures, and voice tone. The physiological signal-based approach collects physiological signals, including Electroencephalography (EEG), Electrocardiography (ECG), Galvanic Skin Response (GSR), Heart Rate Variability (HRV), Respiration Rate (RR), Skin Temperature, and Electromyogram (EMG). In both methods, the collected data is processed using a machine learning pipeline that performs data pre-processing, feature extraction, training, and validating to make predictions \cite{fang2022prevent}. The predicted targets are binary classification tasks derived from 2D or 3D models of emotion. These classifiers split emotion into different dimensions, e.g., arousal, valence, and dominance (if using 3 dimensions) \cite{fang2023introducing}. In the past decade, multiple studies have explored the feasibility of such automated emotion recognition approaches, and several datasets have been published, including ASCERTAIN \cite{subramanian2016ascertain}, MAHNOB-HCI \cite{soleymani2011multimodal}, etc.

\begin{figure*}[ht]
  \centering
  \includegraphics[width=.95\linewidth]{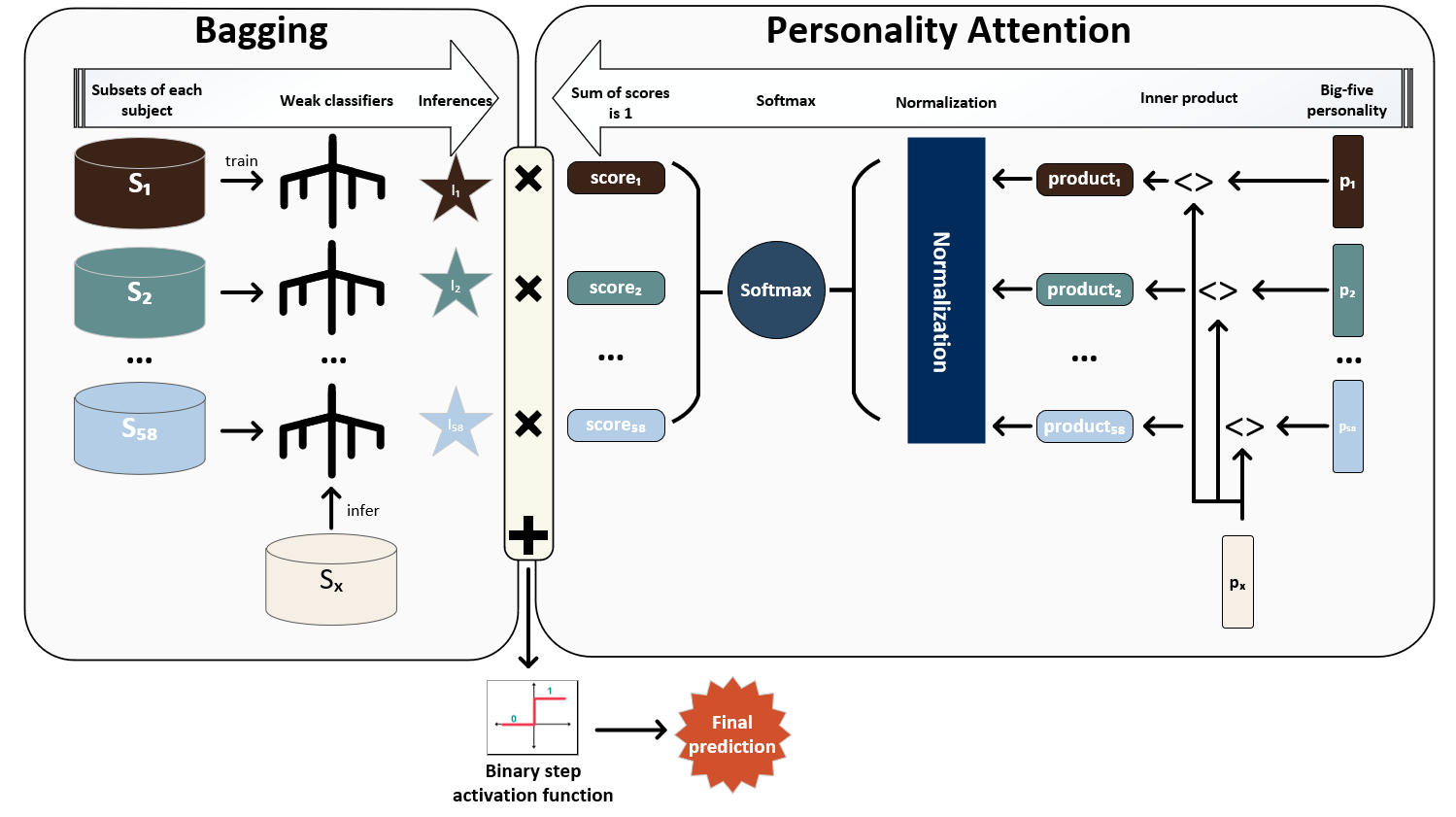}
  \caption{The proposed APEX framework.}
  \label{fig_framework}
\end{figure*}

Recent research has proposed using additional contextual and psychological factors, including personality, in automated emotion recognition. The relationship between personality and effect is first proposed in Eysenck's personality model \cite{eysenck1990biological} and further validated. By using functional magnetic resonance imaging (fMRI), the blood oxygenation level dependent (BOLD) signal was examined to indicate arousal and valence level which further proved the correlation between personality and emotion responses. Different databases have been released for personality-involved automated emotion recognition, including ASCERTAIN \cite{subramanian2016ascertain}. These data sets used emotional videos as stimuli and captured physiological signals while subjects were watching videos. Also, the big-five traits for personality were used to evaluate subjects' personalities. Subramanian et al. \cite{subramanian2016ascertain} deployed traditional machine learning algorithms including support vector machine (SVM) and naive Bayes (NB) \cite{ ouguz2023emotion}, to classify emotions and obtained averaged f1-score of 0.7 for Valence and 0.655 for Arousal. Shao et al. \cite{shao2019emotion} proposed a hypergraph neural network where both a modality and a personality derive hypergraphs, and the vertices in the hypergraphs are subject-video pairs. The hypergraphs output to a fully-connected layer to yield the final prediction. Results on ASCERTAIN dataset showed 80.57\% accuracy in classifying high/low valence and 73.67\% accuracy in classifying high/low arousal. Tian et al. \cite{tian2021personality} first cluster subjects of ASCERTAIN dataset in different groups based on their personality using K-means and applied deep neural network to do classification. However, the former studies have treated personality too lightly or arbitrarily. This study proposes a novel framework that embeds the personality attention mechanism from natural language processing to ensemble learning, specifically, bagging. A personality score is calculated by the inner product of two 5D personality traits. It is used as the reweighing factor of different weak classifiers to yield a more personalized and adapted prediction for the specific subject.

The main contributions of this paper can be summarized as follows: 
\begin{itemize}
    \item We propose a scoring system to calculate the similarity between the personalities of two subjects.
    \item A novel framework that embeds the personality score to ensemble learning is proposed.
    \item We conducted experiments on the ASCERTAIN dataset to verify the feasibility of the proposed framework. Results showed that our proposed method overperforms existing studies.
\end{itemize}

The remainder of this paper is organized as follows. Section \ref{sec_method} introduces the dataset used in the study and provides a detailed description of the proposed approaches. In section \ref{sec_experiments}, we present the experiments conducted and the evaluation results. Section \ref{sec_discussion} discusses this study's merits, potential, weaknesses, and future directions. Lastly, Section \ref{sec_conclusion} concludes this study and summarizes the key learnings.

\section{Method}
\label{sec_method}

The pipeline of the proposed framework is presented in Fig. \ref{fig_framework}. It contains two parts: the bagging part and the personality attention part.

\subsection{Dataset}
We used ASCERTAIN dataset \cite{subramanian2016ascertain} to validate the feasibility of this study. ASCERTAIN dataset is a multimodal dataset for emotion recognition that uses physiological sensors. Fifty-eight subjects were involved in the study and were tested for their big-five personality traits. Subjects were instructed to report their emotional self-ratings of "Arousal," "Valence," "Engagement," "Liking," and "Familiarity" after watching 36 videos that evoke emotion, such as ”The Shining.” Physiological signals were recorded while the subjects watched the videos. The recorded signals included EEG, ECG, GSR, and facial activity. In this study, we adopted the 2D emotion model, which has two dimensions "Arousal" and "Valence." Thus, we designate two classification tasks: arousal and valence. In addition, we adopted two physiological signals, ECG and GSR, because these two signals reveal critical information and can be deployed on wearable devices.

\subsection{APEX framework}
Fig. \ref{fig_framework} shows the APEX framework, which consists of Bagging and Personalized attention. The critical assumption of this framework is that personality and emotional response are correlated; hence participants with similar personalities evoke similar emotional responses. Thus, when doing emotion recognition for a particular participant, we want the training data with a similar personality to contribute more to the classification and reduce the weights of data with less personality similarity. 

To do so, we treat each participant as an independent training sub-set. Assume the total number of participants in the training set is $\mathbf{N}$, then $\mathbf{N}$ training sub-sets are used. Each training sub-set $\mathbf{S_i}$ is used to train a weak classifier and in total, there are $\mathbf{N}$ weak classifiers, where $i$ is the index for a training sub-set. We use decision trees as the weak classifier in this study, but this choice is highly flexible. After all the weak classifiers are trained, they can be used to infer a new subject's data $\mathbf{S_x}$. This will yield $\mathbf{N}$ classification results $\mathbf{I_i}$ in probability form, i.e., a possibility number ranging in $[0,1]$, where $i$ is the index of the weak classifier. 

On the other hand, each training sub-set has five-dimensional personality traits, represented as a five-dimensional data vector $p_i$ and the test subject also has personality traits $p_x$. To calculate the score, we use the inner product of between each training sub-set $p_i$ and the testing subject $p_x$ to yield $\mathbf{N}$ products $product_i$:
\begin{equation}
    {product}_i = \langle \mathbf{p}_i\; , \; \mathbf{p}_x \rangle
\end{equation}

All products are pushed to normalization to re-scale to $[0,1]$. Then, they were transformed to probability scores $\mathbf{score}_i$ by being imported to a softmax function. After softmax function, the sum of $\mathbf{score}_i$ is 1. The scores are calculated as:
\begin{equation}
\begin{aligned}
    score_i = \mathrm{softmax}(product_i) \\
    = \frac{\exp(product_i)}{\sum_{j=1}^m \exp(product_j)} \in \mathbb{R}.
\end{aligned}
\end{equation}

\begin{table*}[hbp]
  \caption{Comparison of APEX and Existing Studies Performance}
  \label{table_results}
  \centering
  \begin{tabular}{cllcc}
    \hline
    \bfseries Study  & \bfseries Model & \bfseries Task & \bfseries Accuracy  & \bfseries AUC\\
    \hline \hline
    \multirow{2}*{Santamaria et al.\cite{santamaria2018using}}  & \multirow{2}*{SVM/NB (no personality)}  & Arousal & 65.5\%  & 0.70\\
    ~  & ~ &Valence & 70.3\%  & 0.73\\
    \hline
    \multirow{2}*{Shao et al.\cite{shao2019emotion}}  & \multirow{2}*{Hypergraph Neural Network}  & Arousal & 70.8\% &  0.77\\
    ~  & ~ &Valence & 72.9\% & 0.79\\
   \hline
    \multirow{2}*{Tian et al.\cite{tian2021personality}}  & \multirow{2}*{K-mean and Deep Neural Networks}  & Arousal & 71.1\% & 0.73\\
    ~  & ~ &Valence & 70.6\%  & 0.73\\
    \hline
    \multirow{2}*{\textbf{APEX Framework}}  & \multirow{2}*{Attention-based Bagging}  & Arousal & \textbf{77.1\%}  & \textbf{0.83}\\
    ~  & ~ &Valence & \textbf{76.9\%}  & \textbf{0.81}\\

  \hline
\end{tabular}

\end{table*}

To embed the personality attention to bagging model, the inferences are multiplied with scores one by one and summed up to yield the final prediction:
\begin{equation}
    Final Prediction = \mathrm{BinaryStep}(\sum_{i=1}^N score_i\times I_i)
\end{equation}
Note that the output inferences are probability values indicating the possibility of the predicted label, and it is a fraction ranging in $[0,1]$. Thus, in order to transform the probability value into a binary classes, a binary step activation function is deployed here, where the function is described as:
\begin{equation}
    \mathrm{BinaryStep}(x) = 
    \begin{cases}
    0,  & \text{if }x < 0.5\\
    1,  & \text{if }x \geq 0.5
    \end{cases}
\end{equation}
In the end, the final prediction is a class which embodies well-weighed decisions from all training sub-sets.

\section{Experiments and Results}
\label{sec_experiments}
To validate the feasibility of the proposed framework, we conducted experiments on the ASCERTAIN dataset. We used ECG and GSR data in this study. In the pre-processing stage, a low-pass filter with a 0.2 Hz cut-off frequency was used to obtain the low-pass GSR signal. Next, a band-pass filter with 0.67 to 40 Hz cut-off frequencies was applied to the ECG signals. Then, a five seconds time window with a five seconds window shift is used to extract features. In total, 42 features are extracted. Feature selection is applied, including variance threshold, SelectKbest, and tree-based selecter, to optimize the feature space. After feature selection, the ten most important features are kept, as shown in Table \ref{table_feature}. 



Then, the data were normalized to the scale of $[0,1]$. This is because we hypothesize that different subjects have different vital signs baselines and have different physiological responses to emotional videos. In order to lessen the bias among subjects, normalization was applied. Afterward, data were imported into the APEX framework. Each subject's feature set becomes a training subset $S_i$ to train a weak classifier. 

We used Decision Tree as the weak classifier as it overperformed other classifiers in our preliminary experiments. Decision tree, support vector machine (SVM), k-NN, linear discriminative analysis have been tested and decision tree overperformed the second best classifiers, SVM by 0.9\% in accuracy. We consider this is due to the advantages of the decision tree algorithm itself, as it can be considered a collection of if-then rules, which are highly interpretable, and fast in prediction. In addition, for a decision tree, there is no need to consider whether the features are interdependent. However, when using the decision tree algorithm alone, it is prone to overfitting. Through various methods, the complexity of the decision tree is suppressed, the fitting ability of a single decision tree is reduced, and multiple decision trees are integrated by the ensemble method, which can solve the problem of overfitting. Thus, the ensemble learning method and the decision tree learning algorithm can complement each other and are a perfect pair. 

\begin{table}[h]
\caption{Feature List}
\label{table_feature}
\begin{tabular}{cll}
\hline
\bfseries Category             & \bfseries Feature  & \bfseries Description                                                          \\ \hline \hline
\multirow{6}{*}{HRV} & MAV      & Mean Absolute Value                                                  \\ \cline{2-3} 
                     & Range    & Range                                                                \\ \cline{2-3} 
                     & SDNN     & Standard deviation of NN intervals                                   \\ \cline{2-3} 
                     & RMSSD    & \makecell[l]{Root mean square of successive\\ RR interval differences}               \\ \cline{2-3} 
                     & pNN50    & \makecell[l]{Percentage of successive RR intervals that\\ differ by more than 50 ms} \\ \cline{2-3} 
                     & TINN     & Baseline width of the RR interval histogram                          \\ \hline
\multirow{4}{*}{GSR} & MAV      & Mean Absolute Value                                                  \\ \cline{2-3} 
                     & P2P      & Peak to Peak Amplitude                                               \\ \cline{2-3} 
                     & VAR      & Variance                                                             \\ \cline{2-3} 
                     & MeanFreq & Mean Frequency                                                       \\ \hline
\end{tabular}
\end{table}
There are 58 participants in the ASCERTAIN dataset. However, due to the quality of the physiological signals, ten subjects with relatively low data quality were removed from the study. Thus, 48 weak classifiers were trained. For testing, we set each subject as the testing set at a time, excluded this particular test subject from the weak classifiers, and used the other 47 weak classifiers to infer the test subject. This process was repeated 48 times for all subjects in the dataset. We reported the averaged system's performance from all subjects.


To assess the effectiveness of our novel framework, we conducted a comprehensive evaluation by replicating three seminal studies that tackled emotion recognition, both with and without the incorporation of personality traits. In contrast to these established works, our proposed framework was implemented and compared against their methodologies. In the study by Santamaria et al. \cite{santamaria2018using}, a convolutional neural network was employed to classify arousal and valence using the AMIGOS dataset. Their deep learning model demonstrated substantial efficacy in emotion recognition through physiological signals in the absence of personality integration. Shao et al. \cite{shao2019emotion} pursued a distinctive approach by employing a hyperedge algorithm to establish connections between personality attributes, physiological signals, and emotions. Their work highlighted the potential synergy between these facets, showcasing the intricate interplay between personality and emotional states. In a similar vein, Tian et al. \cite{tian2021personality} leveraged K-means clustering to categorize participants based on their personality traits. Subsequently, deep neural networks were trained for each personality subgroup. Our framework reproduced and extended their signal processing, feature extraction, and model deployment procedures on the ASCERTAIN dataset, utilizing the same participant cohort. The outcomes of our endeavors are meticulously presented in Table \ref{table_results}, underscoring the advancements brought about by our proposed framework.

\begin{figure}[ht]
  \centering
    \includegraphics[width=0.9\linewidth]{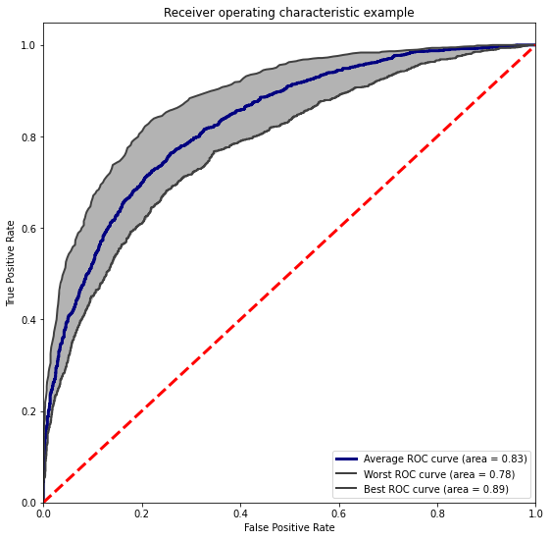}
  \caption{The ROC curve of APEX framework on Arousal task from ASCERTAIN dataset.}
  \label{fig_roc}
\end{figure}

The results showed our proposed framework overperformed existing studies. Our proposed framework achieved 77.1\% and 76.9\% accuracy on arousal and valence tasks, respectively, which overperformed 6\% and 4\% compared with the best results from baseline studies. Fig. \ref{fig_roc} shows the receiver operating characteristic curve of APEX working on the Arousal task from ASCERTAIN dataset. As we have 48 different classifier sets and individual testing sets, 48 ROC curves were produced. We averaged the ROC curves by the y-axis. All ROC curves lay within the gray zone and the boundaries represent the best and worst test cases of the area under curves equal to 0.78 and 0.89, respectively.

\section{Discussion}
\label{sec_discussion}

This study proposes a novel framework that integrates personality score into ensemble learning so that when inferencing, the decision pay more attention to subjects with similar personality. The APEX framework has a core of re-weight, it not only consider the personality similarity to adjust and pay more attention to more corelatted training samples while also keep decisions from physiological signals as dominant. We consider this is the main reason APEX is more accurate when compared with state-of-the-art approaches involving personality in emotion recognition where personality is treated arbitrarily (e.g., cluster subjects into personality groups) or considered slightly (e.g., a modality in parallel with other physiological signal modalities). This idea enables the APEX framework to progress more than the baseline (without personality) on a higher hierarchy.

In the personality attention part, several products are produced after the inner product of two personality vectors. We add a normalization process before sending them to softmax. This is because a personality vector is 5-dimensional with values ranging from $[1,7]$ and the average personality trait value in the ASCERTAIN dataset is 4.7. Thus, the output inner products usually range in hundred scales which is quite a large number when importing to softmax. In such case, the output probability will be concentrated to the highest value, i.e., 0.99 for the highest subject and 0.01 for all rest subjects' summation. Thus, the decision will almost all depend on that specific subject with the 0.99 score. This is not the ideal case, so a normalization process is added to prevent the such phenomenon.

In the proposed study, decision tree is selected as the weak classifier in the framework from four classifiers. However, other models can also be embedded such as linear classifiers, shallow neoral networks, etc. They may have different performances on different problems consider the type of data, the complexity of relationships and the potential challenges and computation time requirments.

To extend this study, other factors that may influence emotion recognition other than personality can be taken into consideration, for example, demographic information. Exploring more factors and finding out the influential group of factors can lead emotion recognition to a more personalized and accurate stage. In addition, other algorithms can be used. This could include modifying the training process of neural networks by letting fewer score subjects only contribute to the first few epochs.

\section{Conclusion}
\label{sec_conclusion}
This study proposes the APEX framework that combines attention mechanism and ensemble learning to perform personalized emotion recognition. In the proposed framework, personality scores are calculated and used to re-weigh the outputs from weak classifiers. We conducted experiments on 48 subjects from ASCERTAIN dataset, and the results showed that the proposed framework performed better than the current state-of-the-art studies. We obtained classification accuracies for valence and arousal were 76.9\% and 77.1\%, respectively.

\bibliography{sample-base}

\end{document}